\documentclass{article}
\usepackage{spconf,amsmath,graphicx}
\usepackage{url}
\usepackage{cite}
\usepackage[hidelinks]{hyperref}
\usepackage{booktabs}
\usepackage{multirow}
\usepackage{paralist}
\usepackage{subfig}

\title{Investigation of Speaker Representation\\for Target-Speaker Speech Processing}

\name{\begin{tabular}{c}Takanori Ashihara$^{1}$, Takafumi Moriya$^{1}$, Shota Horiguchi$^{1}$, Junyi Peng$^{2}$,\\
Tsubasa Ochiai$^{1}$, Marc Delcroix$^{1}$, Kohei Matsuura$^{1}$, Hiroshi Sato$^{1}$\end{tabular}}
\address{
$^1$NTT Corporation, Japan\\
$^2$Brno University of Technology, Czechia
}

\begin{document}
\ninept


\setlength\floatsep{4pt}
\setlength\textfloatsep{6pt}
\setlength\intextsep{6pt}
\setlength\abovecaptionskip{2pt}
\setlength\belowcaptionskip{2pt}
\setlength\abovetopsep{4pt}
\setlength\dblfloatsep{0pt}
\setlength\dbltextfloatsep{4pt}
\aboverulesep=0.25ex 
\belowrulesep=0.5ex 

\maketitle

\begin{abstract}
Target-speaker speech processing (TS) tasks, such as target-speaker automatic speech recognition (TS-ASR), target speech extraction (TSE), and personal voice activity detection (p-VAD), are important for extracting information about a desired speaker's speech even when it is corrupted by interfering speakers.
While most studies have focused on training schemes or system architectures for each specific task, the auxiliary network for embedding target-speaker cues has not been investigated comprehensively in a unified cross-task evaluation.
Therefore, this paper aims to address a fundamental question: what is the preferred speaker embedding for TS tasks?
To this end, for the TS-ASR, TSE, and p-VAD tasks, we compare pre-trained speaker encoders (i.e., self-supervised or speaker recognition models) that compute speaker embeddings from pre-recorded enrollment speech of the target speaker with ideal speaker embeddings derived directly from the target speaker's identity in the form of a one-hot vector.
To further understand the properties of ideal speaker embedding, we optimize it using a gradient-based approach to improve performance on the TS task.
Our analysis reveals that speaker verification performance is somewhat unrelated to TS task performances, the one-hot vector outperforms enrollment-based ones, and the optimal embedding depends on the input mixture.
\end{abstract}
\vspace{0.0cm}

\vspace{-0.1cm}
\section{Introduction}
\label{sec:intro}
\vspace{-0.1cm}
In daily conversational scenarios, a desired (or target) speaker's speech is often degraded by the speech of other (interfering) speakers.
To tackle such realistic situations, several speech processing models have been extended to enable conditioning with cues from a target speaker~\cite{ts_asr1,ts_asr2,ts_asr3,ts_asr4,ts_asr5,tse1,tse2,tse3,tse_comp_emb,tse4,p_vad1,p_vad2,p_vad3,p_vad4}, referred to as target-speaker speech processing (TS) models.
TS models aim to extract information about a target speaker from a speech recording of multiple speakers.
While TS tasks are more challenging than conventional single-speaker tasks due to the need to handle two processes---identifying the speaker of interest and extracting desired information---in a single model, this speaker-conditioned speech processing has attracted much attention, achieving significant improvements in multi-talker speech processing challenges~\cite{chime6_stc,m2met2}.
\par
TS tasks cover various areas depending on the nature of the information to be extracted, such as content, speech, and segments.
For \textit{content}, target-speaker automatic speech recognition (TS-ASR)~\cite{ts_asr1,ts_asr2,ts_asr3,ts_asr4,ts_asr5} focuses on transcribing utterances of a target speaker from a speech mixture and has shown to be more effective than standard ASR models without any speaker cues.
Regarding \textit{speech}, target speech extraction (TSE) is designed to estimate a clean signal of a target speaker from a mixture of talkers~\cite{tse1,tse2,tse3,tse4,tse_comp_emb,ts_asr1}.
With respect to \textit{segments}, personal voice activity detection (p-VAD) focuses on determining whether a frame of audio contains a target speaker's speech in a mixture sequence~\cite{p_vad1,p_vad2,p_vad3,p_vad4}.
\par
\begin{figure*}[t]
  \centering
  \subfloat[][System overview used in each TS task]{\includegraphics[height=5.3cm]{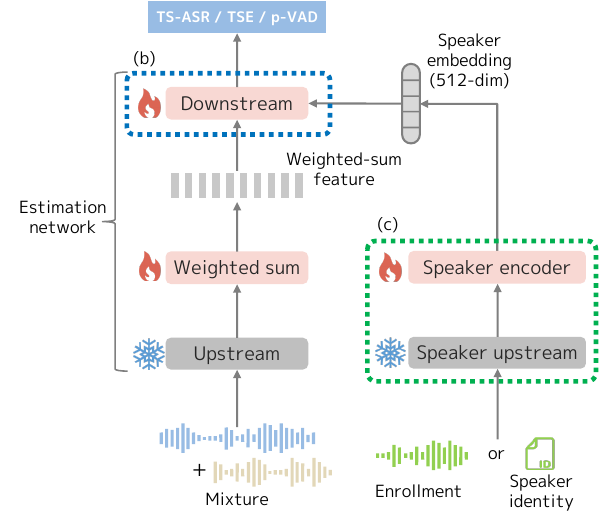}\label{fig:met_a}}
  \hfill
  \subfloat[][Downstream models for each TS task]{\includegraphics[height=5.3cm]{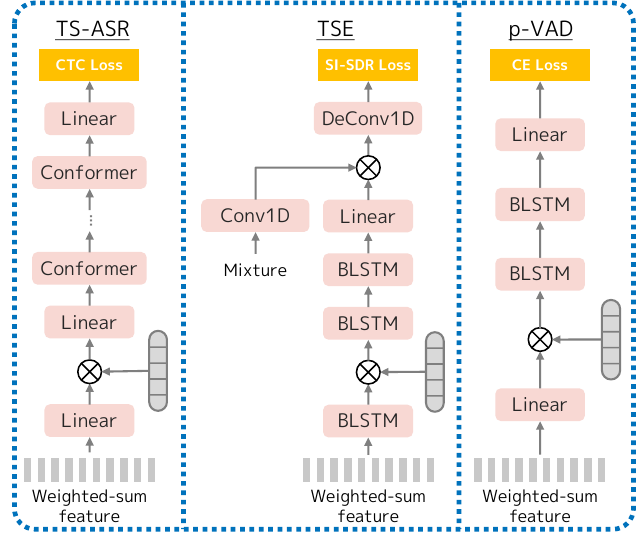}\label{fig:met_b}}
  \hfill
  \subfloat[][Auxiliary network]{\includegraphics[height=5.3cm]{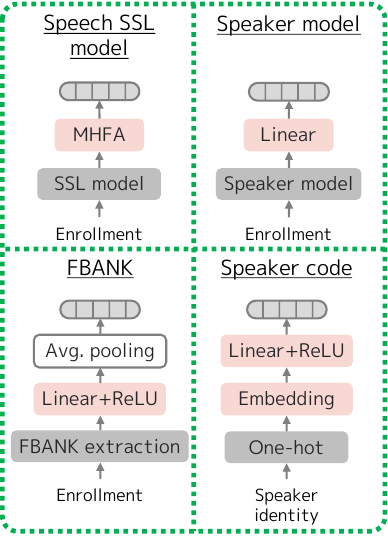}\label{fig:met_c}}
  \caption[]{Schematic diagrams of the SUPERB-based TS evaluation system.}
  \label{fig:met}
\end{figure*}

As shown in Fig.~\ref{fig:met_a}, standard TS systems developed with a neural network (NN) consisting of a recognizing/enhancing network (hereinafter referred to as an estimation network) with a speech mixture as input, which is conditioned on a speaker embedding.
This embedding helps to identify the target speaker in a mixture.
It is computed with an auxiliary network that accepts a pre-recorded enrollment of a target speaker as input.
Several options for obtaining speaker embeddings have been proposed~\cite{ts_asr1,ts_asr3,tse1,tse2,tse3}, such as using signal processing acoustic features or a one-hot vector corresponding to a speaker identity (hereinafter called the \textit{speaker code})~\cite{tse2} passed through multilayer NNs as an auxiliary network.
Other studies employed pre-trained speaker representations obtained from off-the-shelf models such as a Gaussian mixture model (i.e.,~i-vector~\cite{i_vector})~\cite{ts_asr2,tse1,p_vad2}, NN-based supervised speaker models (such as d-vector~\cite{d_vector} and x-vector~\cite{x_vector})~\cite{tse_comp_emb,ts_asr4,p_vad1,p_vad3,p_vad4}, and self-supervised learning (SSL) models~\cite{tse_comp_emb,tse4}.
\par
Despite the progress of TS systems, studies on the properties of auxiliary networks and trained speaker representations are relatively limited.
Therefore, the present study addresses a fundamental question: what auxiliary networks or speaker representations are suitable for TS tasks?
Moreover, by conducting a cross-task evaluation, we aim to understand whether the optimal speaker representation is consistent or varies across different TS tasks.
Answering this question may indicate whether the synergy between the tasks should be explored more thoroughly.
\par
In this paper, we explore speaker embeddings from an auxiliary network by comprehensively comparing the performance of diverse SSL models, supervised speaker recognition models, and a speaker code on TS-ASR, TSE, and p-VAD under a unified experimental environment.
Intuitively, a speaker code applicable only to \textit{speaker-closed} conditions, where the test speaker is fully included in the training set, would surpass enrollment-based models, in which embeddings are derived from the enrollment samples.
This is because a speaker code-based TS system can directly learn embeddings that optimally capture speaker characteristics for each TS task, unlike an enrollment-based system.
However, it is unclear if there are even more optimal speaker embeddings that could further improve TS tasks.
Therefore, inspired by successes in the computer vision community~\cite{cv_invert1,cv_invert2,cv_invert3,cv_invert4}, we additionally perform gradient-based optimization on the speaker embedding so that the score for the true class is maximized.
Our major findings are as follows:
(1) While a different auxiliary network is suitable for each TS task, SSL models, including both typical Transformer-based and ECAPA-TDNN-based models, are generally effective as auxiliary networks for all three tasks.
(2) Although the representation of speaker embeddings tends to form clusters by speaker, especially for the TS-ASR and TSE tasks, the performance of the automatic speaker verification (ASV) task is irrelevant to the performance of the TS tasks.
(3) While a speaker code outperforms other auxiliary networks, gradient-based optimization reveals that the embedding can be further refined.
(4) Gradient-based optimization also shows that the optimal embedding shifts to make one speaker more distinguishable from others and varies according to the input mixture.
We hope that the results of this exploration will provide insight into developing more effective speaker representations for TS tasks.

\vspace{-0.1cm}
\section{Method}
\label{sec:method}
\vspace{-0.1cm}
\subsection{SUPERB-based target-speaker speech processing system}
\label{met:ts-superb}
\vspace{-0.0cm}
To evaluate speaker representations from an auxiliary network, we developed a unified experimental environment across TS tasks.
As illustrated in Fig.~\ref{fig:met_a}, the design of the estimation network follows the SUPERB framework~\cite{superb}.
The input to the downstream model consists of a weighted-sum sequence of the outputs from each layer of the upstream model, where only the weights and downstream model are trainable, while the upstream model parameters remain frozen.
By adopting the SUPERB framework, speaker representation can be evaluated efficiently using downstream models with relatively small parameters compared to training an entire system from scratch.
This efficiency is made possible by speech SSL upstream models pre-trained on large amounts of unlabeled data.
To condition the estimation network on target-speaker cues, the SUPERB framework is extended to accept speaker embeddings.
Here, 512-dimensional speaker embeddings are used for all tasks in this paper.

\vspace{-0.0cm}
\subsubsection{Downstream model in estimation network}
\vspace{-0.0cm}
Figure~\ref{fig:met_b} illustrates the downstream model for each TS task.
Note that we always use the Hadamard product for speaker conditioning in the estimation network regardless of the downstream model and task.\footnote{We experimented with different conditioning methods, such as addition and concatenation, but the Hadamard product proved to be the most effective across all three tasks.}
Furthermore, we empirically determined which layer to condition with speaker embeddings.
\par
\textbf{TS-ASR:}
The left panel of Fig.\ref{fig:met_b} shows the downstream model architecture for TS-ASR.
Although the architecture of the ASR task in SUPERB utilizes two layers of bidirectional long short-term memory (BLSTM) with 1024 dimensions, the TS-ASR system using this architecture exhibited instability of training and unreasonably low performance in preliminary experiments.
Therefore, we employ multiple blocks of Conformer~\cite{conformer}, as utilized in previous ASR/TS-ASR studies\cite{conformer,ts_asr3,ts_asr4}.
Since one of the purposes of this research is to compare auxiliary networks in a unified system rather than achieving state-of-the-art performance in TS-ASR, we empirically chose the Conformer model with the minimum number of parameters (approximately 7 million) while ensuring reasonable performance.
The speaker embedding is applied to the output from the first linear layer, which aligns the dimension of the weighted-sum feature with the embedding (i.e., 512 dimensions).
This conditioned feature is then fed into a subsequent linear layer to match the dimensions of the Conformer blocks.
Finally, an additional linear layer is applied on top of the Conformer blocks to predict a token sequence of the target speaker.
The tokens and training loss adhere to the ASR task in SUPERB, specifically character tokens and connectionist temporal classification (CTC) loss~\cite{ctc}.
\par
\textbf{TSE:}
The middle section of Fig.\ref{fig:met_b} depicts the architecture of the downstream model used for TSE.
The downstream architecture is based on the speech enhancement (SE) task in SUPERB-SG\cite{superb_sg} with some modifications.
Specifically, similar to the original SE task in SUPERB-SG, the system also consists of three BLSTM layers to predict the mask for the target speaker's signal.
To mask the mixture signal, the SE recipe in SUPERB-SG utilizes short-time Fourier transform (STFT) and inverse STFT (iSTFT) to encode and decode the mixture signal and masked spectral signal, respectively.
Meanwhile, a recent SSL-based TSE system~\cite{tse4} demonstrated higher extraction performance compared to the STFT/iSTFT pipeline by utilizing 1D-convolution and deconvolution layers.
This improvement is attributed to the joint optimization of the encoder/decoder, which led to task-suitable frequency bands.
Therefore, we also incorporate 1D-convolution and deconvolution layers for the encoder/decoder.
To inject the target speaker cue, the speaker embedding is multiplied by the output from the first BLSTM layer, which is determined empirically.
For the training objective, we minimize the scale-invariant signal-to-distortion ratio (SI-SDR)\cite{si_sdr} between the inferred masked signal and the target speaker's clean signal.
According to a previous study~\cite{tse4}, this objective achieves better extraction performance than the mean square error used in the SE recipe of SUPERB-SG.
\par
\textbf{p-VAD:}
In Fig.\ref{fig:met_b}, the right panel illustrates the downstream model architecture employed for p-VAD.
We use the \textit{embedding conditioned training} from the original p-VAD paper\cite{p_vad1} with certain adjustments to align with the other TS tasks.
Specifically, the weighted-sum feature is fed into a linear layer to match the dimension of the speaker embedding, and the features are multiplied by the embedding using the Hadamard product.
The speaker-conditioned feature is then processed by two BLSTM layers, followed by a linear layer that predicts three classes: non-speech (\texttt{ns}), target speaker speech (\texttt{tss}), and non-target speaker speech (\texttt{ntss}) using the BLSTM output.
To train the model, we minimize the cross-entropy (CE) loss as described in the original p-VAD paper~\cite{p_vad1}.

\vspace{-0.0cm}
\subsubsection{Speaker upstream and encoder models in auxiliary network}
\label{met:spk-enc}
\vspace{-0.0cm}
Figure~\ref{fig:met_c} illustrates the auxiliary networks consisting of a speaker upstream and a speaker encoder.
For the speaker upstream, similar to the upstream and downstream models of SUPERB, the parameters are always fixed to ensure a fair comparison of pre-trained models.
In contrast, the parameters of the speaker encoder are updated during training in each TS task so that task-optimized embeddings can be established.
\par
\textbf{Speech SSL model:}
To encode an enrollment waveform, our study applies speech SSL models as a speaker upstream, as displayed in the top left of Fig.~\ref{fig:met_c}.
Regarding the speaker encoder, since the speech SSL model's output is a sequential feature, the sequence needs to be accumulated over time to form a single embedding vector.
We used multi-head factorized attention (MHFA)~\cite{mhfa} to aggregate layer- and frame-wise features, which has been shown to be more effective than temporal averaging of the weighted-sum features.
\par
\textbf{Speaker model:}
We can also use speaker embedding extractors trained for speaker recognition tasks (e.g., speaker identification (SID) and ASV)~\cite{i_vector,d_vector,x_vector,ecapa_tdnn,spkssl_mse,spkssl_simclr_moco,spkssl_vicreg,spkssl_dino_er,spkssl_c3dino,spkssl_dino_compara,spkssl_ca_dino,spkssl_rdino} as an auxiliary network.
For example, the network architecture is time-delay neural networks (TDNNs) used by x-vector~\cite{x_vector} or ECAPA-TDNN~\cite{ecapa_tdnn}.
Our study employs the speaker models as one of the variations of an auxiliary network, as displayed in the top right of Fig.~\ref{fig:met_c}.
The output vector is further fed into a linear layer to align the dimensions to 512.
\par
\textbf{FBANK:}
Several TS models have used hand-crafted acoustic features as input instead of outputs from pre-trained models~\cite{ts_asr1,ts_asr3,tse1,tse2,tse3}.
The FBANK auxiliary network used in our study is depicted in the bottom left of Fig.~\ref{fig:met_c}.
Specifically, log mel-filterbank features extracted from an enrollment speech are fed into a linear layer with a rectified linear unit (ReLU), followed by temporal average pooling to aggregate information in the time domain.
\par
\textbf{Speaker code:}
In a \textit{speaker-closed} task setting, where the target speaker of a test set is completely seen during training, a one-hot vector corresponding to the speaker's identity (called speaker code) could be a suitable speaker embedding, as it can optimally represent speaker properties for TS tasks.
Therefore, we apply a speaker code to the TS tasks as illustrated in the bottom right of Fig.~\ref{fig:met_c}.
For the architecture, a one-hot vector corresponding to a speaker identity is transformed by an embedding layer and input into a linear layer with ReLU.
While auxiliary networks other than a speaker code can work with a \textit{speaker-open} condition where test speakers are not seen during training, the speaker code is only applicable in a \textit{speaker-closed} condition, where test speakers are fully duplicated in the training data.
Although this limitation is somewhat impractical, we verify its performance and representation as one of the most effective auxiliary networks.

\vspace{-0.0cm}
\subsection{Gradient-based speaker embedding optimization}
\label{met:grad_opt}
\vspace{-0.0cm}
\begin{figure}[t]
  \centering
  \includegraphics[width=8.5cm]{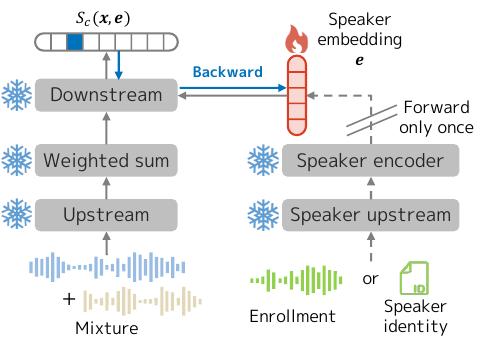}
  \caption{Schematic diagram of gradient-based speaker embedding optimization. The forwarding process of an auxiliary network is executed only once to obtain an initial speaker embedding to be optimized.}
  \label{fig:met_bwd}
\end{figure}
To explore optimal speaker embeddings, we propose directly optimizing a speaker embedding using a gradient calculated from a mixture and the corresponding ground truth of a target speaker.
This approach is summarized in Fig.\ref{fig:met_bwd} and is inspired by the successes of prior computer vision studies~\cite{cv_invert1,cv_invert2,cv_invert3,cv_invert4}.
\par
In the procedure, using a mixture and enrollment speech or speaker identity, weighted-sum feature $\mathbf{x}_t$ and speaker embedding $\mathbf{e}_0$ are calculated from an upstream model followed by weighted-sum computation with trained weights and a trained auxiliary network, respectively.
These features are then fed into a trained downstream model to obtain a score $S_{c_t}(\mathbf{x}_t,\mathbf{e}_0)$ (i.e.,~an output of a final prediction head before a softmax function) of a true class $c_t$.
Here, a true class $c_t=[c_1,\dots,c_T]$ is, for example, a token sequence corresponding to a ground-truth transcription for the TS-ASR task, where $t$ and $T$ are a temporal frame index and the total number of frames, respectively.
To optimize the embedding directly, we use the unnormalized class score to discover an embedding where $S_{c_t}$ achieves a high value (i.e.,~$\arg\max_{\mathbf{e}} S_c(\mathbf{x},\mathbf{e})$).
This has been shown to be more suitable than using the posteriors for analyzing the network behavior, as the class posterior can be maximized by minimizing the scores of other classes~\cite{cv_invert1,cv_invert2}.
Note that while we attempted to minimize the standard CE loss in preliminary TS-ASR experiments, we also verified that optimizing the embedding by maximizing the score led to a more significant reduction in error rates compared to the slight reduction achieved through CE loss minimization.
During optimization, we iteratively subtract a gradient multiplied by a step size $\alpha$ from an input embedding to adjust it in a direction that increases the score.
The update equation at the $n$-th iteration is
\begin{equation}
\mathbf{e}_{n+1} = \mathbf{e}_{n} - \alpha \nabla_{\mathrm{e}} \sum\nolimits_t S_{c_t}(\mathbf{x}_t,\mathbf{e}_{n}),
\label{eq:iteration}
\end{equation}
where $n\in\{0,\dots,N-1\}$.
Here, $\nabla_{\mathrm{e}} \sum\nolimits_t S_{c_t}(\mathbf{x}_t,\mathbf{e}_{n})$ and $N$ denotes the gradient of a score with respect to an embedding and the number of iterations for optimization, respectively.
During optimization, the speaker embedding is trainable, while the other parameters remain frozen.
In this paper, we apply this approach to the trained TS-ASR models, as TSE is not a classification problem, making it difficult to straightforwardly apply score maximization.
Additionally, the performance of p-VAD has almost reached saturation, leaving seemingly little room for further optimization.

\vspace{-0.2cm}
\section{Experimental setup}
\label{sec:setup}
\vspace{-0.2cm}
\subsection{Target-speaker speech processing system}
\vspace{-0.2cm}
The system for the TS tasks was based on S3PRL\footnote{\url{https://github.com/s3prl/s3prl}} developed for SUPERB~\cite{superb}.
While SUPERB mainly focuses on comparing the performance of upstream models, our objective is to compare auxiliary networks.
Therefore, we utilized WavLM~{\sc Base+}~\cite{wavlm} as the upstream model, since it is reported to be the optimal upstream model in TSE~\cite{tse4}.
Note that we observed a similar trend with the HuBERT {\sc Base}, although the overall performance was generally lower.
\par
\textbf{Speech SSL model:}
For the speaker upstream model, we compare wav2vec2.0 {\sc Base} \cite{wav2vec2}, HuBERT {\sc Base} \cite{hubert}, data2vec {\sc Base} \cite{data2vec}, WavLM {\sc Base}, WavLM {\sc Base+}, and WavLM {\sc Large} \cite{wavlm}, as summarized in Table~\ref{tab:result_wlm}.
All models are publicly available (or directly callable via S3PRL).
After extracting features through speaker upstream models, MHFA, configured with eight attention heads and a compression layer with a dimension of 128, was applied to yield an embedding with 512 dimensions.
\par
\textbf{Speaker model:}
We employed x-vector~\cite{x_vector} and ECAPA-TDNN~\cite{ecapa_tdnn}, which were trained using the WeSpeaker toolkit~\cite{wespeaker}.
These models were pre-trained on supervised speaker identification tasks with angular additive margin softmax loss, followed by large margin fine-tuning, a technique widely used in speaker verification challenges~\cite{voxsrc2022}.
Other training details followed the WeSpeaker's VoxCeleb2~\cite{voxceleb} recipes.
The architectures were identical to that of the originals~\cite{x_vector,ecapa_tdnn}, with the ECAPA-TDNN models having different numbers of channels (512 and 1024, denoted as ECAPA-TDNN-c512 and ECAPA-TDNN-c1024, respectively).
Additionally, considering recent developments in utterance-wise speech SSL models for speaker representation~\cite{spkssl_mse,spkssl_simclr_moco,spkssl_vicreg,spkssl_dino_er,spkssl_c3dino,spkssl_dino_compara,spkssl_ca_dino,spkssl_rdino,ashihara2024}, we employed the \textit{self-DIstillation with NO labels} (DINO)~\cite{dino_org} model with the ECAPA-TDNN architecture and 1024 channels.
The DINO model was trained in an SSL fashion using a self-distillation framework that transfers knowledge from a teacher model to a student model, resulting in higher ASV performance than other utterance-wise speech SSL models (for more details, see the previous papers~\cite{dino_org,spkssl_dino_er,spkssl_c3dino,spkssl_dino_compara,spkssl_ca_dino,spkssl_rdino,ashihara2024}).
The training configuration for the DINO model is identical to the WeSpeaker setting.\footnote{\url{https://github.com/wenet-e2e/wespeaker/tree/master/examples/voxceleb/v3/dino}}
Note that the amount of pre-training data from VoxCeleb2 differs from that of the off-the-shelf speech SSL models as described above.
Nevertheless, VoxCeleb2 is widely used in ASV tasks and competitions~\cite{voxsrc2022} and contains a relatively large amount of data with supervisory speaker labels.
Therefore, we believe it is a reasonable choice for comparing the most promising and widely accepted speaker models in recent years, similar to the speech SSL models, offering valuable insights into aspects such as model efficiency and effectiveness relative to data and model size.
\par
\textbf{FBANK:}
We extracted 80-dimensional log mel-filterbank outputs using a 25-ms window and a 10-ms shift.
The linear layer that processes the acoustic features has 512 dimensions.
\par
\textbf{Speaker code:}
The input speaker identity depends on the number of speakers present in the training set.
As described in Section~\ref{ssec:task}, we utilized the training and evaluation dataset from Libri2Mix, which includes 251 speakers.
The one-hot vector was transformed into a 512-dimensional embedding through an embedding layer, followed by a linear layer with 512 dimensions and a ReLU activation.
Note that although we conducted comprehensive preliminary experiments without the linear layer and ReLU (i.e.,~using only the embedding layer), the TS models did not converge.

\begin{table*}[h!]
\caption{Evaluation results of each task with various network combinations. Each slash-separated cell of the TS tasks shows the results on the \texttt{clean} subset (left) and the \texttt{both} subset (right) in Libri2Mix. The ASV task follows SUPERB~\cite{superb} except for downstream model architectures. \texttt{LS}, \texttt{LL}, \texttt{GS}, \texttt{VP}, and \texttt{VC} denote LibriSpeech~\cite{librispeech}, Libri-Light~\cite{librilight}, GigaSpeech~\cite{gigaspeech}, VoxPopuli~\cite{voxpopuli}, and VoxCeleb2~\cite{voxceleb}, respectively.}
\label{tab:result_wlm}
\centering
\resizebox{\linewidth}{!}{%
\begin{tabular}{@{}llcl|cccccc@{}}
\toprule
 & & & & \multicolumn{1}{c}{TS-ASR} & \multicolumn{3}{c}{TSE} & \multicolumn{1}{c}{p-VAD} & \multicolumn{1}{c}{ASV} \\ \cmidrule(lr){5-5} \cmidrule(lr){6-8} \cmidrule(lr){9-9} \cmidrule(l){10-10}
 Auxiliary network & Speaker upstream & Model size & Pre-training data & WER (\%)$\downarrow$ & SI-SDR (dB)$\uparrow$ & STOI (\%)$\uparrow$ & PESQ$\uparrow$ & mAP (\%)$\uparrow$ & EER (\%)$\downarrow$ \\
\midrule
\midrule
FBANK & FBANK & - & - & 27.62 / 50.34 & 8.62 / 7.16 & 84.60 / 76.77 & 1.782 / 1.377 & \textbf{99.00} / 97.31 & 25.72 \\
\midrule
\multirow{6}{*}{Speech SSL model} & wav2vec2.0 {\sc Base} \cite{wav2vec2} & 94M & \texttt{LS} & 19.25 / 39.53 & 10.50 / 8.40 & 88.68 / 79.58 & \textbf{1.920} / 1.418 & 98.83 / 97.44 & 3.37 \\
 & HuBERT {\sc Base} \cite{hubert} & 94M & \texttt{LS} & 19.08 / 38.74 & 10.57 / 8.39 & 88.66 / 79.64 & 1.912 / 1.415 & 98.93 / 97.42 & 3.08 \\
 & data2vec {\sc Base} \cite{data2vec} & 94M & \texttt{LS} & 19.25 / 38.66 & 10.60 / 8.51 & 88.79 / 79.75 & 1.915 / 1.418 & 98.53 / 96.96 & 3.94 \\
 & WavLM {\sc Base} \cite{wavlm} & 94M & \texttt{LS} & 19.86 / 39.19 & \textbf{10.61} / 8.70 & \textbf{88.85} / 80.30 & 1.917 / \textbf{1.425} & 98.77 / 97.43 & 2.86 \\
 & WavLM {\sc Base+} \cite{wavlm} & 94M & \texttt{LL} + \texttt{GS} + \texttt{VP} & 19.21 / 38.70 & 10.58 / \textbf{8.78} & 88.78 / \textbf{80.50} & 1.907 / 1.419 & 98.37 / 97.15 & 2.10 \\
 & WavLM {\sc Large} \cite{wavlm} & 315M & \texttt{LL} + \texttt{GS} + \texttt{VP} & 19.00 / 37.93 & 10.37 / 8.53 & 88.23 / 79.84 & 1.881 / 1.414 & 98.76 / 97.52 & 1.79 \\
\midrule
\multirow{4}{*}{Speaker model} & x-vector \cite{x_vector} & 5M & \texttt{VC} & 19.01 / 40.33 & 10.01 / 7.80 & 87.54 / 78.26 & 1.856 / 1.384 & 98.76 / 96.95 & 2.50 \\
 & ECAPA-TDNN-c512 \cite{ecapa_tdnn} & 6M & \texttt{VC} & 19.68 / 39.74 & 9.89 / 7.81 & 87.42 / 78.33 & 1.851 / 1.392 & 98.50 / 96.59 & 1.73 \\
 & ECAPA-TDNN-c1024 \cite{ecapa_tdnn} & 15M & \texttt{VC} & 20.64 / 40.76 & 9.64 / 7.65 & 86.99 / 77.94 & 1.856 / 1.383 & 98.28 / 96.70 & \textbf{1.32} \\
 & ECAPA-TDNN-DINO & 15M & \texttt{VC} & \textbf{18.84} / \textbf{37.88} & 10.59 / 8.40 & \textbf{88.85} / 79.63 & 1.904 / 1.401 & 98.82 / \textbf{97.54} & 2.67 \\
\bottomrule
\end{tabular}%
}
\end{table*}

\vspace{-0.2cm}
\subsection{Downstream tasks}
\label{ssec:task}
\vspace{-0.2cm}
\textbf{Dataset:}
We utilized LibriMix~\cite{librimix} for the training and evaluation dataset of all TS tasks.
Specifically, the training set was the \texttt{train-100} subsets in Libri2Mix (i.e.,~two speakers in a mixture), which included \texttt{clean} and \texttt{both} subsets (without any noise and with WHAM! noise~\cite{wham}, respectively).
The corresponding enrollment speech was prepared according to SpeakerBeam's configuration\footnote{\url{https://github.com/BUTSpeechFIT/speakerbeam}}~\cite{tse1,tse3}.
Following this configuration, an enrollment speech utterance was assigned on a one-to-one basis for each mixture in the test and development subsets, unlike in the training set, where it was selected randomly from multiple enrollment utterances of the target speaker.
The test subsets prepared by the above procedure were called the \textit{speaker-open} evaluation set and were used unless otherwise mentioned.
For the speaker code experiment, as described in Section~\ref{met:spk-enc}, the speakers in the test data must be included in the training data.
Therefore, we created an additional test subset called the \textit{speaker-closed} set, consisting of 6000 mixtures, which is the same number as the \textit{speaker-open} set.
The speakers of this subset were fully contained in the training set, while the utterances were not duplicated.
All datasets were sampled at 16kHz.
For the true label of the p-VAD task, we created labels from the diarization labels\footnote{\url{https://github.com/s3prl/LibriMix}} used in the speaker diarization task of SUPERB.
\par
\textbf{Task configuration:}
The training hyperparameters were based on the ASR\footnote{\url{https://github.com/s3prl/s3prl/tree/main/s3prl/downstream/asr}}, SE\footnote{\url{https://github.com/s3prl/s3prl/tree/main/s3prl/downstream/separation_stft2}}, and speaker diarization\footnote{\url{https://github.com/s3prl/s3prl/tree/main/s3prl/downstream/diarization}} recipes, for the TS-ASR, TSE, and p-VAD tasks, respectively.
Only changes from the original configuration are described below.
\par
For \textbf{TS-ASR}, the learning rate was warmed up to 0.001 in the first 15k steps and then linearly decayed for the remaining steps.
Eight Conformer blocks were used, each consisting of four attention heads with 144 dimensions, a 1D convolution with a kernel size of 15, and a feedforward module with 1024 dimensions.
Performance was reported in terms of word error rates (WERs).
\par
For \textbf{TSE}, the total training and validation steps were set to 150k and 4k steps, respectively.
The Adam optimizer with 3k warm-up steps followed by linear decay was used.
The evaluation metrics were SI-SDR~\cite{si_sdr}, short-time objective intelligibility (STOI)~\cite{stoi}, and perceptual evaluation of speech quality (PESQ)~\cite{pesq}.
\par
For \textbf{p-VAD}, two BLSTM layers with 128 cells were used as a downstream model.
The trained models were evaluated by calculating the mean average precision (mAP) over all classes.
\par
In addition to the TS tasks described above, the performance of \textbf{ASV} was measured to evaluate the quality of speaker representation.
While we followed the ASV setup of SUPERB using VoxCeleb1~\cite{voxceleb}, the downstream models were substituted to align with the auxiliary network.
Specifically, we utilized the MHFA for speech SSL models and identity processing for speaker models.
Identity processing is a parameter-less module that treats the output of speaker models as the resultant embedding and has been reported to outperform the original downstream models (i.e.,~multilayer TDNN)~\cite{ashihara2024}.
The models are evaluated in terms of equal error rates (EERs).

\vspace{-0.2cm}
\subsection{Gradient-based speaker embedding optimization}
\vspace{-0.2cm}
The number of iterations $N$ explained in Section~\ref{fig:met_bwd} was set to 100, which was large enough to saturate performance.
We utilized the Adam optimizer to determine the direction of the gradient, i.e., $\nabla_{\mathrm{e}} \sum\nolimits_t S_{c_t}(\mathbf{x}_t,\mathbf{e}_{n})$ in \autoref{eq:iteration}.
Since the original dataset does not have frame-level transcriptions, we used outputs from the standard single-speaker ASR model as the pseudo-ground-truth labels.
The ASR model was trained using Libri2Mix without interfering speakers and demonstrated a 3.78\% WER on the \textit{speaker-closed} subset.
The learning rate was empirically optimized for each auxiliary network and was set to 4 for WavLM {\sc Base+} and 1 for FBANK and the speaker code.

\vspace{-0.4cm}
\section{Results}
\vspace{-0.4cm}
\label{sec:result}
\subsection{Task performances}
\label{ssec:res_per}
\vspace{-0.2cm}
Table~\ref{tab:result_wlm} shows the results for each TS task using the WavLM {\sc Base+} upstream model on the \texttt{clean} and \texttt{both} subsets, i.e., without any noise and with WHAM! noise, respectively.
FBANK demonstrated the lowest performance for all tasks except for p-VAD, suggesting that the pre-training scheme is promising for the auxiliary network of all TS tasks.
However, the FBANK result in p-VAD suggested that simple signal processing methods may be sufficient for relatively easier tasks.
Notably, while experimenting with a deeper speaker encoder, FBANK performance improved up to two linear layers but remained poor.\footnote{For example, the FBANK model with two linear layers achieved 23.92\% and nearly reached saturation on the \texttt{clean} set for the TS-ASR task.}
As for pre-trained models, although the supervised ECAPA-TDNN models (ECAPA-TDNN-c512 and ECAPA-TDNN-c1024) achieved higher ASV performance than the speech SSL models, their performance on the TS tasks was comparable or lower.
This suggests that \textit{the performance between TS and ASV tasks is uncorrelated}, indicating the need for other metrics or tasks to identify a suitable pre-trained speaker upstream for an auxiliary network rather than relying on ASV, which is a basic task evaluating the quality of speaker representation.
This tendency extends the findings of previous research~\cite{tse_comp_emb}, which only validated the TSE task, to the TS-ASR and p-VAD tasks.
Moreover, comparing the optimal configuration across the tasks, we observed that \textit{ECAPA-TDNN-DINO was the most effective for all three tasks}.
This result suggests that the synergy between the TS tasks could be exploited to obtain more suitable speaker representations for speaker-conditioned speech processing.

\par
Table~\ref{tab:result_spkcode} demonstrates the results of the speaker code on the \textit{speaker-closed} evaluation subset under two conditions, \texttt{clean} and \texttt{both}.
For all three tasks, the speaker code demonstrates higher performance in both conditions compared to the other models.
Therefore, we can conclude that \textit{the speaker code provides one of the most effective speaker representations for all three tasks}, although the \textit{speaker-closed} condition is not practical as described in Section~\ref{met:spk-enc}.
Future work includes investigating methods to minimize the distance between the representations of pre-trained models and the speaker code, for example, through knowledge distillation.

\begin{table}[t]
\caption{Evaluation results for speaker code on the \textit{speaker-closed} evaluation sets. WavLM {\sc Base+} is used as the upstream model.}
\label{tab:result_spkcode}
\centering
\resizebox{\linewidth}{!}{%
\begin{tabular}{@{}ll|ccc@{}}
\toprule
 & & \multicolumn{1}{c}{TS-ASR} & \multicolumn{1}{c}{TSE} & \multicolumn{1}{c}{p-VAD} \\ \cmidrule(lr){3-3} \cmidrule(lr){4-4} \cmidrule(l){5-5} 
Mixture & Speaker upstream& WER (\%)$\downarrow$ & SI-SDR (dB)$\uparrow$ & mAP (\%)$\uparrow$ \\
\midrule
\midrule
\multirow{3}{*}{\texttt{clean}} & FBANK & 32.81 & 9.90 & 98.96 \\
 & WavLM {\sc Base+} \cite{wavlm} & 21.79 & 11.31 & 98.87 \\
 & Speaker code & \textbf{18.56} & \textbf{11.70} & \textbf{99.82} \\
\midrule
\multirow{3}{*}{\texttt{both}} & FBANK & 55.29 & 8.02 & 97.57 \\
 & WavLM {\sc Base+} \cite{wavlm} & 43.18 & 9.21 & 97.49 \\
 & Speaker code & \textbf{37.50} & \textbf{9.60} & \textbf{98.95} \\
\bottomrule
\end{tabular}%
}
\end{table}

\vspace{-0.25cm}
\subsection{Embedding of each model and task}
\vspace{-0.25cm}

\begin{figure*}[t]
  \centering
  \subfloat[][Speaker-open evaluation set]{\includegraphics[height=6.7cm]{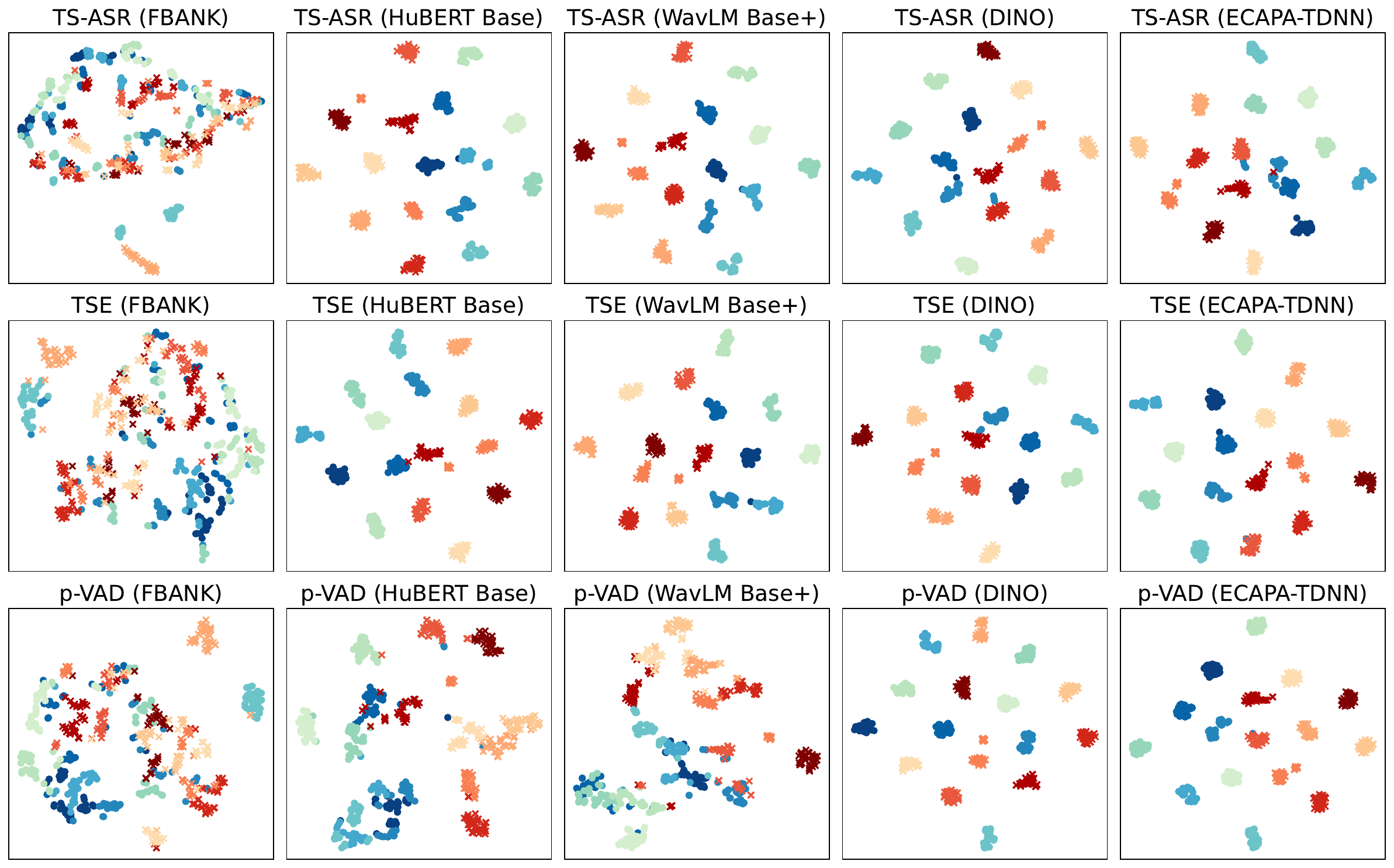}\label{fig:res_fwd_emb_a}}
  \hfill
  \subfloat[][Speaker-closed evaluation set]{\includegraphics[height=6.7cm]{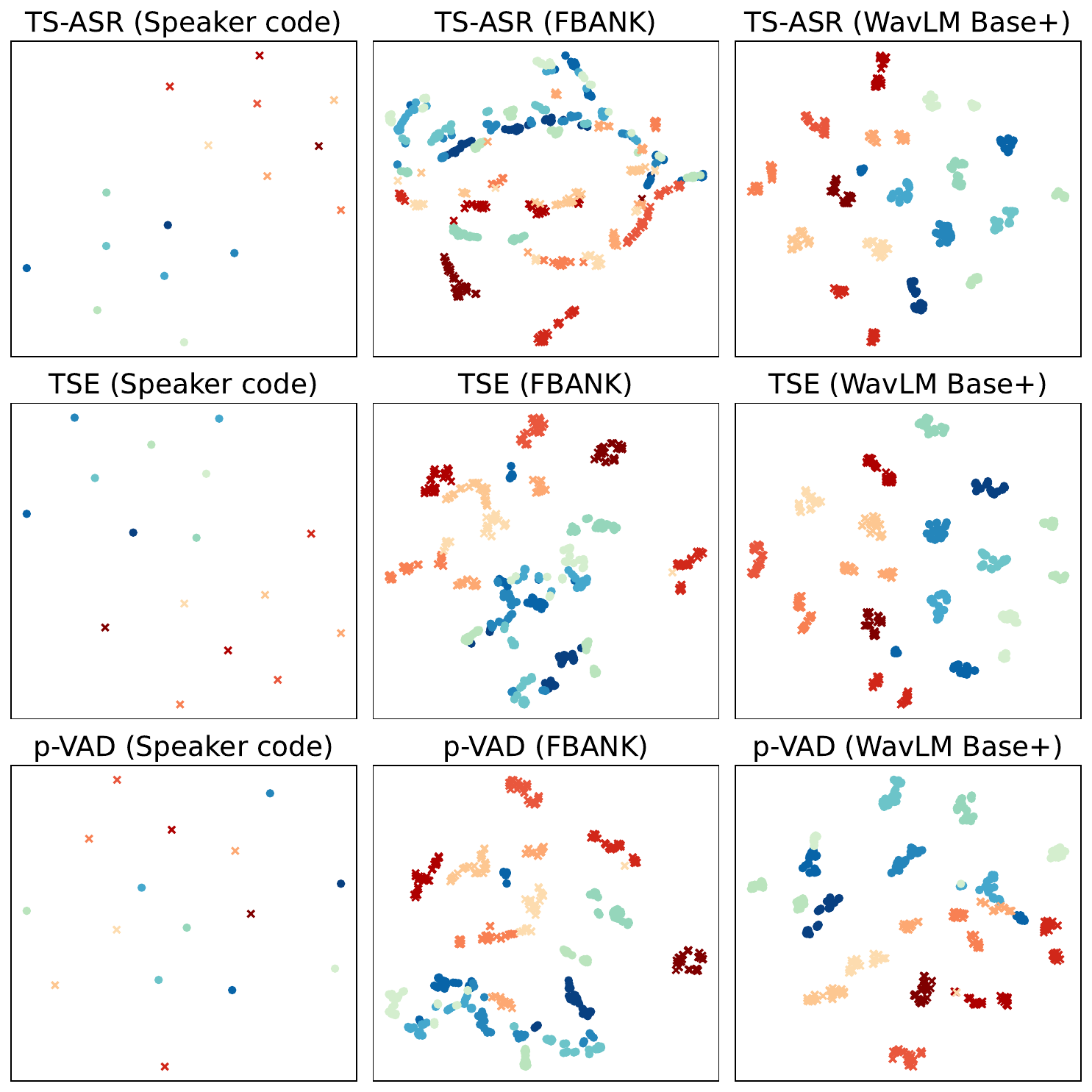}\label{fig:res_fwd_emb_b}}
  \caption[]{Visualization results of speaker embedding. Each color represents each speaker. Blue square and red cross markers indicate male and female speakers, respectively. DINO and ECAPA-TDNN denote ECAPA-TDNN-DINO and ECAPA-TDNN-c1024 models.}
  \label{fig:res_fwd_emb}
\end{figure*}
\par
Figure~\ref{fig:res_fwd_emb} visualizes 512-dimensional speaker embeddings reduced to two dimensions using t-distributed stochastic neighbor embedding (t-SNE)~\cite{t_sne}.
The panels in each column represent each auxiliary network, and the panels in each row represent each TS task.
As shown in Fig.~\ref{fig:res_fwd_emb_a}, which shows the embeddings on the \textit{speaker-open} set, \textit{the pre-trained models (i.e., the right four columns) form speaker clusters, especially for the TS-ASR and TSE tasks}.
Note that the speech SSL models were not explicitly trained to group by speaker, indicating the importance of representing the same speaker similarly for TS tasks.
However, the panels in the last row exhibit high intra-speaker variability, suggesting that p-VAD does not require strict speaker representation that provides similar embeddings for the same speaker.
This is supported by the fact that, while clear speaker clusters appear even in p-VAD, the performance of ECAPA-TDNN models is comparable to or lower than that of other models.
Figure~\ref{fig:res_fwd_emb_b}, which illustrates the embeddings on the \textit{speaker-closed} subset, shows a similar tendency for FBANK and WavLM {\sc Base+}.
The embeddings of the speaker code tend to be grouped by gender, except for the p-VAD task, which further supports the notion that strictly grouped embeddings are not necessary for the p-VAD task.

\vspace{-0.25cm}
\subsection{Embedding of gradient-based optimization}
\vspace{-0.25cm}

\begin{figure*}[h!]
  \centering
  \subfloat[][Evaluation performances]{\includegraphics[height=6.0cm]{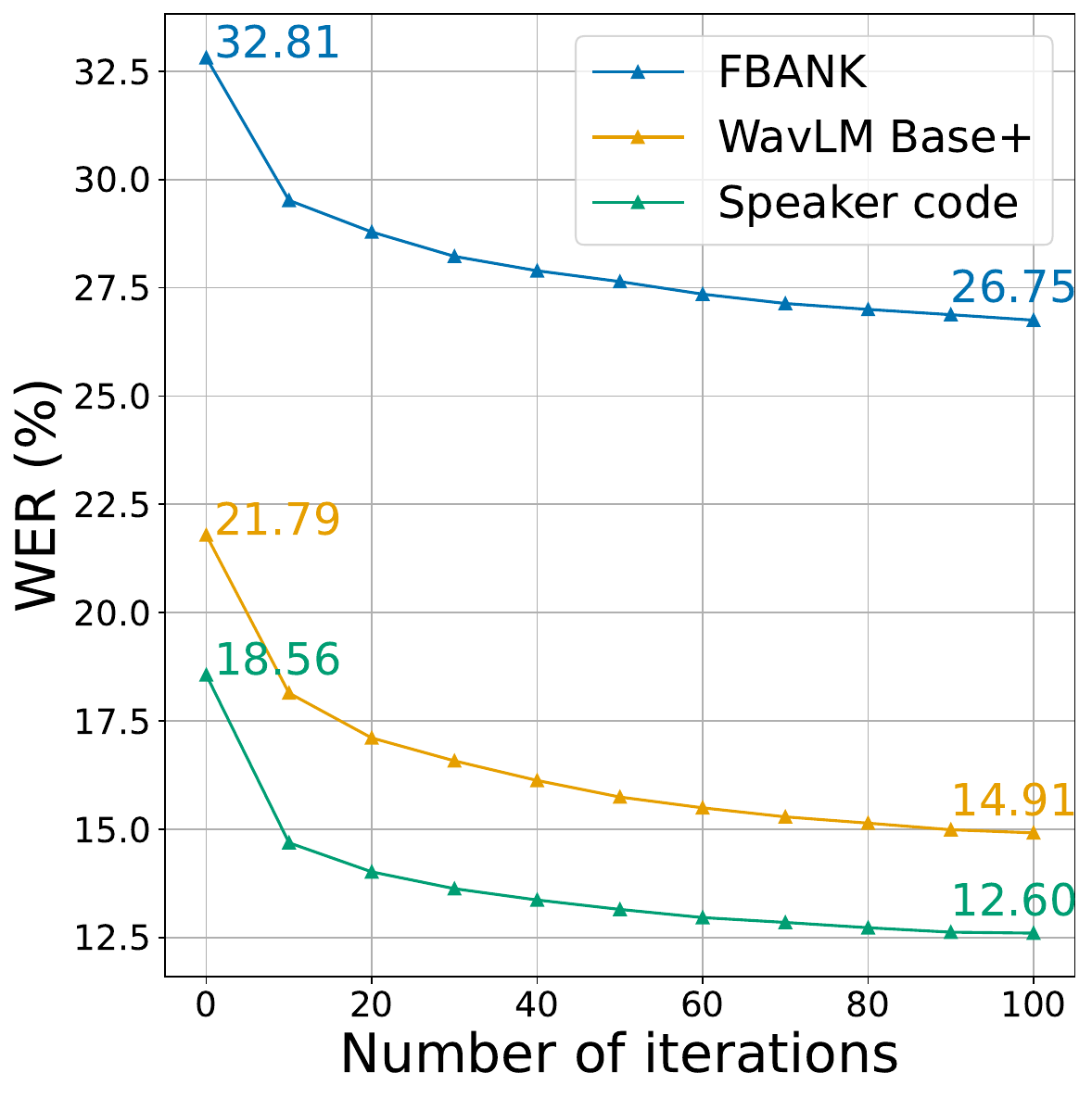}\label{fig:res_bwd_performance}}
  \hspace{1.0cm}
  \subfloat[][Visualization]{\includegraphics[height=6.3cm]{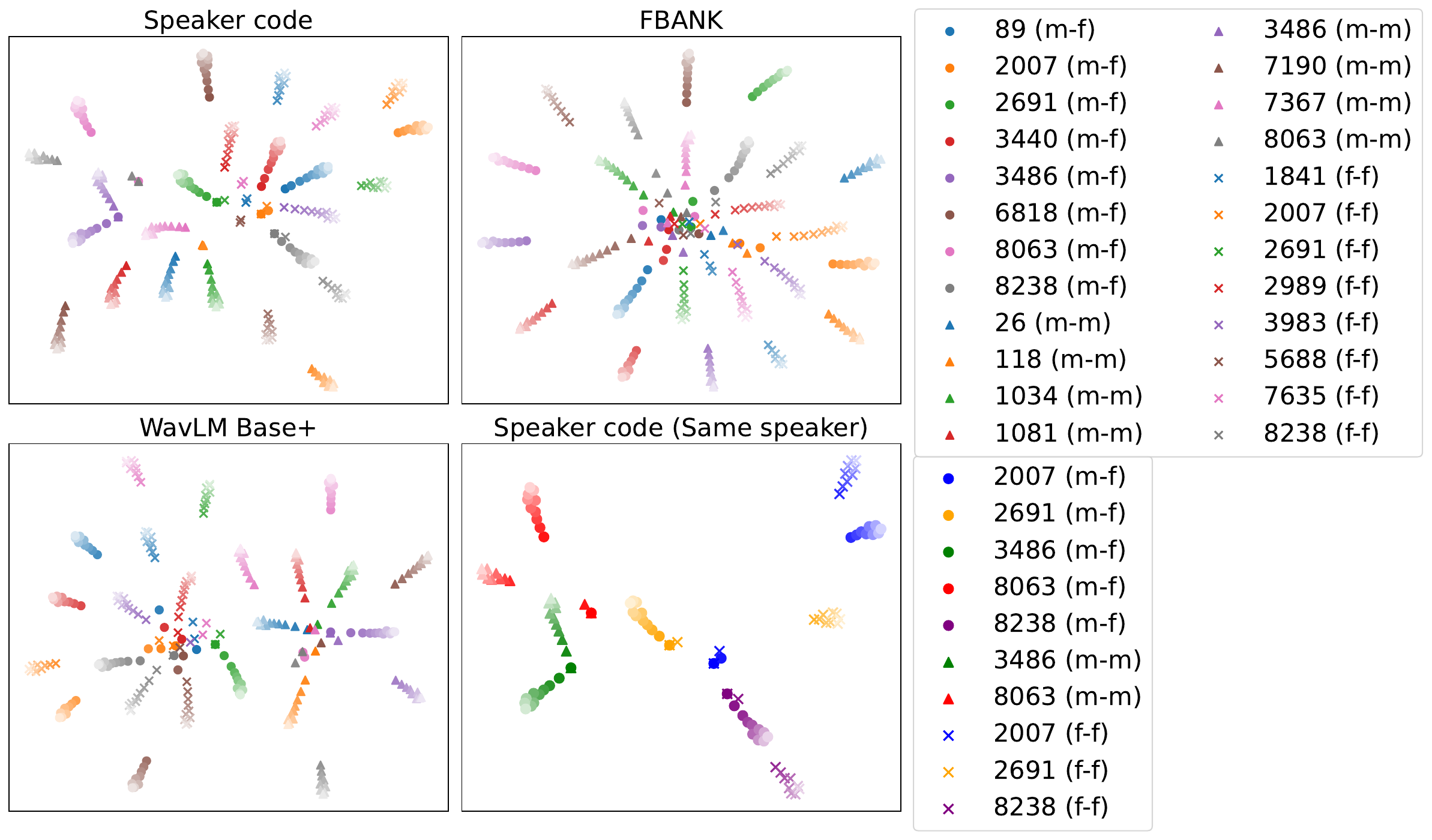}\label{fig:res_bwd_emb}}
  \caption[]{Gradient-based optimization results. (a)Evaluation results using optimized embeddings on TS-ASR at each number of iterations. (b)Visualization results of speaker embeddings at each iteration for each speaker on the TS-ASR task. Each color represents each speaker. The colors are depicted to fade from darker to lighter as the number of iterations increases. Different markers represent different mixture conditions (\texttt{m-f}, \texttt{m-m}, and \texttt{f-f} indicating mixtures of male and female, male and male, and female and female, respectively).}
  \label{fig:bwd_res}
\end{figure*}

The speaker code is one of the optimal auxiliary networks, as described in Section~\ref{ssec:res_per}.
To explore further improvement, we apply gradient-based optimization to the speaker embedding.
Figure~\ref{fig:res_bwd_performance} shows the WERs for each iteration when only the speaker embedding is iteratively optimized to maximize the score of the true class.
Note that since this experiment was conducted on the \texttt{clean} mixture, the performance at the 0-th iteration is the same as in Table~\ref{tab:result_spkcode}.
Performance improves with each iteration, achieving a relative WER reduction of 32.11\% for the speaker code and 31.57\% for WavLM {\sc Base+}.
This suggests that \textit{refining speaker representation can significantly enhance performance further, even for the speaker code.}
\par
Figure~\ref{fig:res_bwd_emb} visualizes the embedding at iterations from 0 to 100 in increments of 10.
Each panel represents an auxiliary network (speaker code, FBANK, and WavLM {\sc Base+}).
To highlight the representation of the same speaker in two different mixtures, the bottom right panel also visualizes the embeddings of the speaker code for the selected speakers.
\textit{The optimized embedding generally tends to increase the inter-speaker difference.}
Interestingly, \textit{the optimal speaker embedding seems to vary for different mixtures, even for the same target speaker.}
Therefore, future work should develop a system architecture that adjusts speaker embedding according to mixtures.
Note that although we fix all network parameters and only optimize the embedding to try to capture the speaker representation as much as possible, the optimal embeddings may reflect both speaker and content information since they are optimized against the score of the true token class.

\vspace{-0.4cm}
\section{Conclusion}
\label{sec:conclusion}
\vspace{-0.35cm}
This paper investigates the speaker representation for TS tasks, i.e., TS-ASR, TSE, and p-VAD, through performance comparisons and visualization of embeddings.
Our results suggest that SSL models with Transformer-based and ECAPA-TDNN-based architectures are more effective for all three TS tasks than supervised speaker models which demonstrate the highest performance in the ASV task.
While the speaker code-based auxiliary network outperforms other enrollment-based auxiliary networks, gradient-based embedding optimization demonstrates that there is room to further optimize speaker representation.
We hope that these results will contribute to the development of more effective TS systems and speaker embeddings.

\newpage
\bibliographystyle{IEEEbib}
\bibliography{strings,refs}

\end{document}